\documentclass[11pt, draftcls, journal, onecolumn]{IEEEtran} 
\usepackage[symbol]{footmisc}
\usepackage{graphics}
\usepackage{amsmath}
\usepackage{amsfonts}
\usepackage{amssymb}
\usepackage{setspace}
\usepackage{float}
\usepackage{graphicx}
\usepackage{amsthm}
\usepackage{color}
\usepackage{cite}
\usepackage{bm}
\usepackage{psfrag}
\usepackage{multirow}
\usepackage[yyyymmdd,hhmmss]{datetime} % gives time of the compiling
\usepackage{epsfig}
\usepackage{psfrag}
\usepackage{pst-node}
\usepackage{fancyhdr,bbm}
\usepackage{color}
\usepackage{makecell}
\usepackage{wrapfig}
\usepackage{comment}
\usepackage{booktabs}
\usepackage{rotating}

\newcommand{\mulrows}[1]{%
  \begin{tabular}{@{}l@{}}\strut#1\strut\end{tabular}%
}

\usepackage[columnwise,switch,mathlines]{lineno} % this adds line numbers for proof reading 
\newcommand{\be}{\begin{equation}}
\newcommand{\ee}{\end{equation}}

\allowdisplaybreaks

\newlength{\figureheight}
\newlength{\figurewidth}
\definecolor{temporalgreen}{RGB}{0,128,0}

\definecolor{BLUE}{rgb}{0,0,1}

\newdateformat{monthyeardate}{\monthname[\THEMONTH] \THEDAY, \THEYEAR} % specifies the format of the date

\newcommand{\paperTitle}{Space-based Global Maritime Surveillance. Part I: Satellite Technologies}

\floatstyle{boxed}
\newfloat{floatbox}{thp}{lob}[section]
\floatname{floatbox}{Text Box}

\excludecomment{comment}

\begin{document}

\renewcommand{\baselinestretch}{1.4}\small\normalsize

\title{\paperTitle}

\author{Giovanni~Soldi,
		Domenico~Gaglione,~\IEEEmembership{Member,~IEEE},
		Nicola~Forti,
		Alessio~Di~Simone,~\IEEEmembership{Member,~IEEE},
		Filippo~Cristian~Daffin\`{a},
		Gianfausto~Bottini,
		Dino~Quattrociocchi,
		Leonardo~M.~Millefiori,~\IEEEmembership{Member,~IEEE},
		Paolo~Braca,~\IEEEmembership{Senior Member,~IEEE},
		Sandro~Carniel,
		Peter~Willett,~\IEEEmembership{Fellow,~IEEE},
		Antonio~Iodice,~\IEEEmembership{Senior Member,~IEEE},
		Daniele~Riccio,~\IEEEmembership{Fellow,~IEEE}, and 
		Alfonso~Farina,~\IEEEmembership{Life Fellow,~IEEE}\vspace{-8mm}\thanks{G.\ Soldi, D.\ Gaglione, N.\ Forti, L.\ M.\ Millefiori, P.\ Braca and S.\ Carniel are with the NATO Centre for Maritime Research and Experimentation (CMRE), La~Spezia, Italy (e-mail: giovanni.soldi@cmre.nato.int, domenico.gaglione@cmre.nato.int, nicola.forti@cmre.nato.int, leonardo.millefiori@cmre.nato.int, paolo.braca@cmre.nato.int, sandro.carniel@cmre.nato.int). A.\ Di Simone, A.\ Iodice and D.\ Riccio are with the Department of Electrical Engineering and Information Technology (DIETI) of the university of Naples Federico II, Italy (e-mail: alessio.disimone@unina.it, antonio.iodice@unina.it, daniele.riccio@unina.it). F.\ C.\ Daffin\`{a}, G.\ Bottini and D.\ Quattrociocchi are with E-GEOS in Rome, Italy (e-mail: filippo.daffina@e-geos.it, gianfausto.bottini@e-geos.it, dino.quattrociocchi@e-geos.it). 
P.\ Willett is with the Department of Electrical and Computer Engineering, University of Connecticut, Storrs, CT 06269-2157 USA (e-mail: peter.willett@uconn.edu). P.\ Willett was supported in part by NIUVT and by AFOSR under contract FA9500-18-1-0463. A.\ Farina is a professional consultant in Rome, Italy (e-mail: alfonso.farina@outlook.it).
This work was supported in part by the NATO Allied Command Transformation under project SAC000A08.
%by the European Research Council (ERC) under grant 700478 (project RANGER) within the Horizon 2020 program,
%% the European UnionÕs Horizon 2020 research and innovation programme
%% within project RANGER (grant agreement no. 700478),
%% \bl{Project Data Knowledge Operational Effectiveness (DKOE) - SAC000808,} 
%by the Austrian Science Fund (FWF) under grants J3886-N31 and P27370-N30, and by the Czech Science Foundation (GA\v{C}R) under grant 17-19638S.
} 
		}
		
\maketitle
\begin{abstract}
\label{abstract}
Maritime surveillance (MS) is crucial for search and rescue operations,
fishery monitoring, pollution control, law enforcement, migration monitoring, and national security policies.
Since the early days of seafaring, MS has been a critical task for providing security in human coexistence. Several
generations of sensors providing detailed maritime information have become available for large offshore areas in
real time: maritime radar sensors in the 1950s and the automatic identification system (AIS) in the 1990s among
them. However, ground-based maritime radars and AIS data do not always provide a comprehensive and seamless
coverage of the entire maritime space. Therefore, the exploitation of space-based sensor technologies installed on
satellites orbiting around the Earth, such as satellite AIS data, synthetic aperture radar, optical sensors, and global navigation satellite systems reflectometry,
becomes crucial for MS and to complement the existing terrestrial technologies. In the first part of this work, we provide an overview of the main available space-based sensors technologies and 
present the advantages and limitations of each technology in the scope of MS. 
The second part, related to artificial intelligence, signal processing and data fusion techniques, is provided in a companion paper, titled: ``Space-based Global Maritime Surveillance. Part II: Artificial Intelligence and Data Fusion Techniques''~\cite{SpaceAESM_2:J20}.
\end{abstract} 
%Furthermore,
%the increasing availability of information from space-based sensors requires the development of sophisticated
%artificial intelligence and information fusion algorithms. The main scope of this article is to provide a survey
%on the principal space-based sensor technologies available nowadays for MSA. Moreover, we present an overview
%on the most promising artificial intelligence and statistical techniques to extract valuable knowledge for MSA by
%fusing information from heterogeneous sensors. 

\vspace{-2mm}
\section{Introduction}
\label{sec:introduction}
Maritime situational awareness (MSA) is defined as ``an enabling capability which seeks to deliver the required information superiority in the 
maritime environment to achieve a common understanding of the maritime situation, in order to increase effectiveness in the planning 
and conduct of operations'' (14 January 2008, Military Committee endorsed NATO concept for MSA).
 %is the discipline of ``knowing what is happening at sea.'' 
A key component of MSA is maritime surveillance (MS), which is crucial for, e.g., search and rescue operations, fishery monitoring, pollution control, law enforcement, migration monitoring, and 
national security policies 
%% %as undertaken by many institutions, agencies, and bodies  
\cite{GiompapaGFGCD09_24}. 
Recent technological advances in the field of MS allow providing seamless wide-area operational pictures in coastal areas and the oceans, also in real time.  
Indeed, multiple heterogeneous sensors and information sources are nowadays available for MS---all with their advantages and limitations---and a significant research effort has been made for their combination. A major source of information related to commercial ship traffic is the automatic 
identification system (AIS), which is designed to identify and track, using a 
network of ground-based stations, ships with 300 or more gross tonnage (GT) and all passenger ships regardless of their size. However, because not all ships share their position and motion information, the AIS alone is not suitable 
to provide seamless MS. Moreover, AIS data relay via ground-based or nearby ships is almost impossible in remote ocean areas. 
To complement the AIS, another important source of information
%%% for MSA
is represented by
%% are 
%% S-band or X-band 
terrestrial radars 
%% that are 
installed along the coastline. 
%%% These radar sensors are capable of detecting and tracking also ships that aim to conceal their identity. 
%However, their coverage area is limited by line-of-sight propagation 
% (even  though anomalous propagation increases coverage), and they 
%may impact the environment when they transmit with a high power. 
However, their
coverage is limited by line-of-sight propagation~\cite{RicSchHol:B10}, and there may be environmental concerns around their ubiquitous use due to the high radiated power.

\begin{figure*}[!t]
\centering
\includegraphics[width=0.95\textwidth]{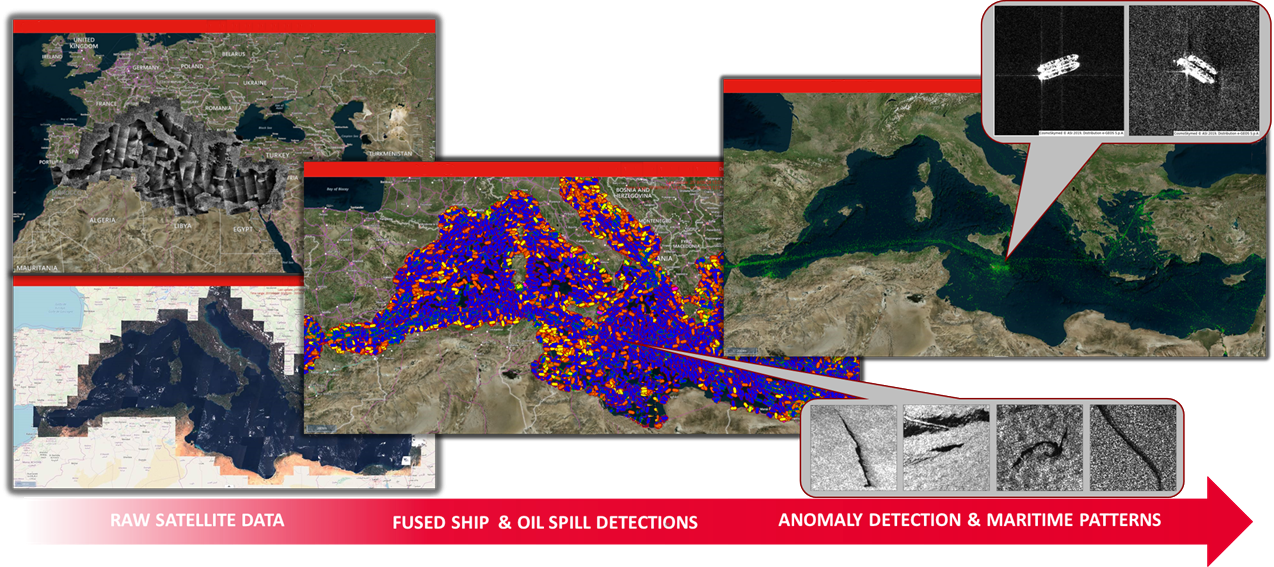}
\caption{The processing of images acquired by space-based sensors with state-of-the-art artificial intelligence and data fusion techniques  to extract information on the current maritime situation enables to detect and prevent piracy, human trafficking, and environmental threats such as oil spills.}
\label{fig:FromRawToAnomalyDetection}
\end{figure*}

Therefore, there is the need of supplementary sensors and technologies to overcome the aforementioned limitations, and provide MS capabilities in remote areas of the Earth. In this context, space-based sensor technologies enable persistent monitoring of 
the maritime domain and ship traffic on a global scale. 
Nowadays, space-based remote sensing data involves a range of technologies and modalities, including satellite AIS (Sat-AIS), 
synthetic aperture radar (SAR), multi-spectral (MSP) and hyper-spectral (HSP) optical sensors, global navigation satellite 
system reflectometry (GNSS-R), with its features, e.g., 
resolutions, viewing angle, frequency, acquisition modes, polarization.  
Space-based sensors for Earth observation (EO) installed on satellites allow collecting images of very large and remote areas 
of the globe with relatively short latency, and hence they are highly relevant
to MS.
Recently, many initiatives from national and international space agencies have encouraged the exploitation of remote sensing 
technologies by providing end-users with freely available data.
The joint NASA/USGS Landsat series, the constellation of satellites for the Mediterranean basin observation (COSMO-SkyMed) commissioned by the Italian Space Agency (ASI), the joint European Union (EU) - European Space Agency (ESA) Copernicus program with the Sentinel multisensor constellation for EO 
and monitoring, the NASA CYGNSS mission for the hurricane forecasts and ocean surface analysis using GNSS signals of opportunity, are only a partial list of space programs and missions providing satellite data at no cost.
The major maritime applications enabled by satellite technologies include, but are not limited to, 
harbor dynamic surveillance, real-time ship surveillance and detection/tracking/classification, sea traffic monitoring, detection of 
security threats and illegal activities and maritime ecosystem management/preservation.

As a consequence of the deployment and spread of space-based sensor technologies, advanced data
processing paradigms, e.g., big data analysis, machine learning, artificial intelligence and data fusion techniques, can now achieve unprecedented 
performance and benefit from the availability of very large data sets. 
In particular, the development of future MS systems combining multiple sensors, terrestrial and space-based, with other sources of information requires dedicated algorithms for the processing of satellite images, detection and 
classification of ships. These algorithms can support MS by processing and organizing the increasing amount of heterogeneous 
information. 
The extracted data and readily-understandable information digested therefrom will help end-users, such as governmental 
authorities, defence forces, coast guards,
and police, to detect anomalies, threats such as oil spills, piracy and human trafficking, and act in time to prevent 
accidents and tragedies.
This paper is divided into two parts: the first part provides an overview of the main available space-based sensors technologies, i.e., Sat-AIS, SAR, MSP and HSP optical sensors, and GNSS-R and 
presents the advantages and limitations of each technology in the scope of MS. The second part of this work~\cite{SpaceAESM_2:J20} focuses on artificial intelligence and data fusion techniques to extract information from raw satellite data and fuse this information to extract valuable and actionable knowledge for end-users and decision-makers.

\section{Satellite Technologies for Maritime Surveillance}
\label{sec:satellite_tech_maritime_surveillance}
Space-based sensors are usually installed on satellites which orbit around the Earth and they can serve different 
purposes. They are used for EO, e.g., SAR and optical sensors; for telecommunication services, e.g., the Italian System 
for Secure Communications and Alerts (SICRAL), the Access on Theatres for European Nation Allied forces - French Italian Dual Use 
Satellite (ATHENA-FIDUS), and the Iridium satellite constellation; and, for navigation purposes, e.g., Galileo, GPS, and GLONASS.
In this work, the focus will be on space-based sensors for EO in the scope of MS. 
Most of the satellites equipped with sensors for EO are placed on a low Earth orbit (LEO), characterized by an 
altitude between 160 km and 2000 km. A polar orbit is a particular LEO in which a satellite passes above or nearly above both 
poles of the Earth on each revolution. 
%Sensors mounted on satellites in polar orbits are able  to acquire the same area of interest multiple times a day above the poles, and at least twice a day above the Equator.
The portion of the Earth surface that a space-based sensor sees while moving along its orbit is referred to as its  
\textit{swath}. Typical satellite swaths in LEO range from tens to hundreds of kilometres, and for this reason they can only observe and communicate with a limited region 
of the Earth at a time. This means that constellations of satellites are needed to provide continuous coverage and guarantee 
a lower revisit time, that is the time interval between two consecutive times that a sensor observes the same region. Sensors mounted on constellations of at least four satellites in polar orbits, e.g., COSMO-SkyMed, are able to illuminate the same area of interest multiple times a day above the poles, and at least twice a day above the Equator. Satellites in 
lower regions of LEO suffer from fast orbital decay, requiring either periodic rebooting to maintain a stable orbit, or 
launching replacement satellites when old ones re-enter. 
Space-based sensors for EO require the acquired images to be transferred to specific ground stations as soon as they are in their field of view. Therefore, the density and the geographical distribution of the ground stations have a direct impact on the latency at which the acquired images are available to the end-user. In order to reduce the latency of the communication with ground stations, constellations of interconnected satellites can be used as tactical assets to transfer the information acquired by satellites to the terrestrial stations almost in real-time. In this context, the European Data Relay System (EDRS), a ESA project within the Advanced Research in Telecommunications Systems (ARTES) program, uses two geostationary data relay satellites to provide links to satellites in LEO, thus enabling real-time communications between spacecraft and terrestrial stations.

In the following sections, we provide a comprehensive overview on the main current space-based sensor technologies that can be exploited for MS. For each technology, we present the main advantages and limitations, when used in the scope of MS. 

\section{Sat-AIS}
\label{sec:overview_satAIS}
In 2002, the International Maritime Organization (IMO) Safety of Life at Sea (SOLAS) convention~\cite{UN_solas} included a mandate that requested many commercial vessels to fit onboard AIS, an anti-collision
broadcast system of transponders automatically exchanging ship traffic information for maritime safety. According to the International Association of Maritime Aids to Navigation and Lighthouse Authorities (IALA), the scope of AIS is ``to improve the maritime safety and efficiency of navigation, safety of life at sea and the protection of the marine environment'' \cite{iala2003technical}. IMO requires ships over 300 GT, cargo vessels over 500 GT, and all passenger ships and all fishing vessels over 45 meters (in EU countries over 15 meters~\cite{EC2011}) to be equipped with an AIS transponder onboard.
The AIS was originally conceived as a short-range identification and tracking system and, when it was designed, it was not anticipated that AIS signals could be received from space. However, satellite AIS receivers have been operational since 2008, thanks to the efforts of several entities that had been experimenting the reception of AIS messages from space.
AIS messages convey information about ship identifier, i.e., the maritime mobile service identity (MMSI), route (position, speed, course, and true heading), and other ship and voyage
characteristics, e.g., cargo type, ship type, speed and maneuvering status, size, destination, and estimated time of arrival.
AIS messages are sent every 2-10 seconds depending of ship type, and the system is capable of handling over 2,000 reports per minute.
AIS signals are broadcast at very high frequency (VHF) bands and, therefore, exhibit a range which is similar to other VHF applications and is mainly dependent on the transmitted power and the range to the local horizon from the VHF antenna, which in turn depends upon the antenna height. Typically, the detection range is 20 and 40 nautical miles for ship-to-ship and ship-to-shore communications, respectively, and therefore can be received and processed only from ships moving close enough to the land-based stations~\cite{hoye2008space,eriksen2006maritime}.

\begin{figure}[!t]
\centering
\includegraphics[width=0.5\columnwidth]{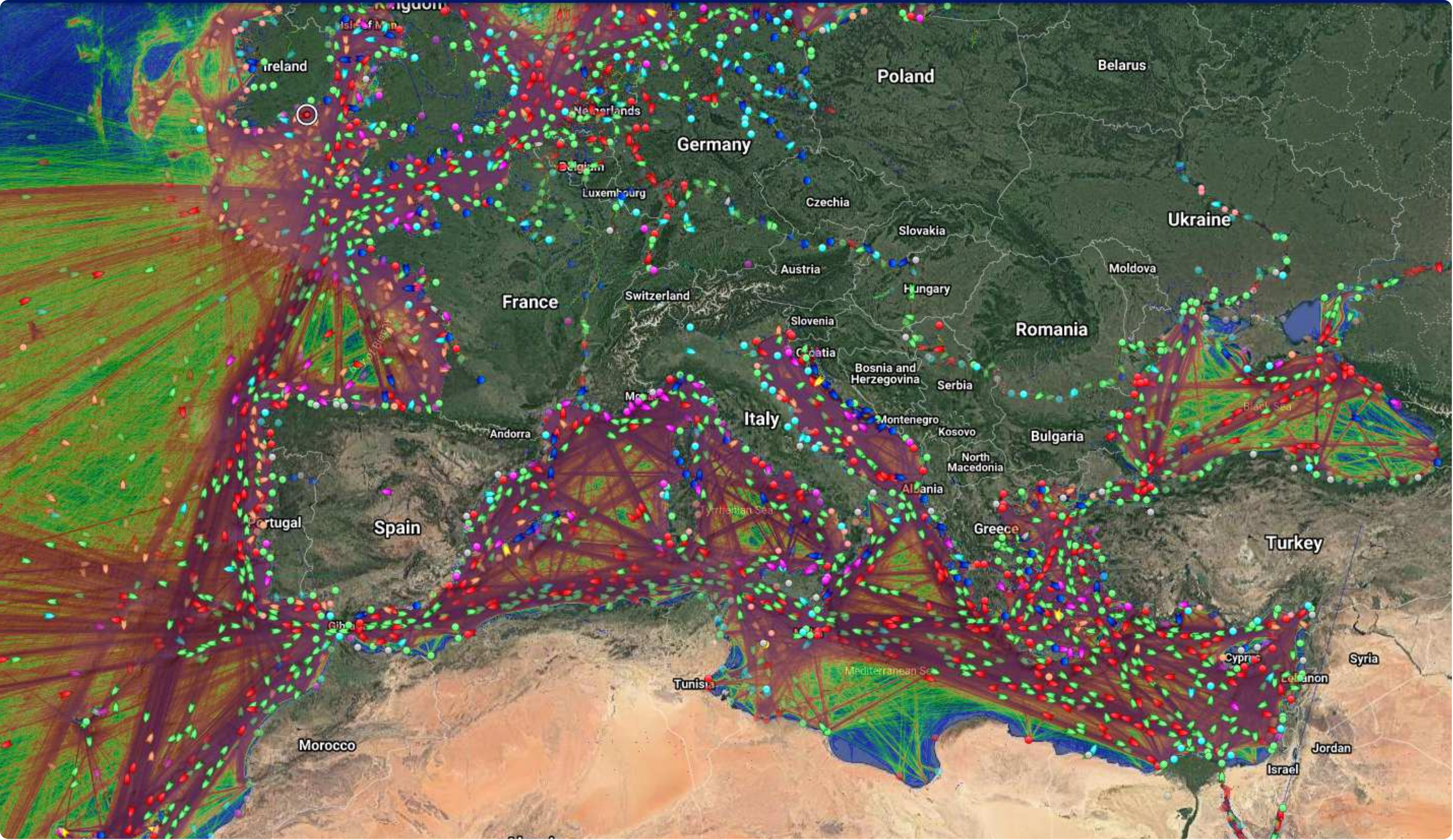}
\caption{Real-time AIS reports in the Mediterranean Sea, acquired by ground-based stations and satellites (image courtesy of MarineTraffic).}
\label{fig:AIS_traffic}
\vspace{-1mm}
\end{figure}
Despite the range improvement with respect to coastal vessel traffic services (VTS), the low coverage of terrestrial AIS remains a crucial limitation as it prevents, at least in principle, the exploitation of AIS services in open sea environments. In particular, the terrestrial reception of AIS signals requires line-of-sight and hence is limited by the curvature of the Earth. This means that ship-to-base and ship-to-ship AIS information is only available around coastal zones or among proximate vessels. This issue is overcome by Sat-AIS systems which are satellites mounting AIS transponders onboard, thus allowing reception of AIS signals within the field of view (FOV) of the receiving satellite antenna. Accordingly, Sat-AIS systems increase extremely the detection range of traditional land-based stations and enable the exploitation of AIS worldwide. Indeed, a satellite AIS transponder in a 600 km altitude orbit exhibits a range to the horizon as large as 1440 nautical miles, giving a key support for global MS and large-area ocean surveillance at marginal cost \cite{eriksen2006maritime}.
Figure~\ref{fig:AIS_traffic} shows real-time AIS data in the Mediterranean Sea, acquired by ground-based stations and satellites.

The feasibility of AIS signal reception at spaceborne altitudes was first discussed in \cite{wahl2005new,thomas2003maritime} and further assessed by Norwegian Defence Research Establishment (FFI) for wide-area maritime surveillance in two European scenarios~\cite{eriksen2006maritime}. The analyses demonstrated a ship detection probability close to 100\% for 1000 vessels within the coverage area, and that reception of AIS signals can achieve power levels 10-20 dB excess above the sensitivity limitations  of standard AIS receivers up to 1000 km \cite{eriksen2006maritime}. 
The first demonstrations  were designed and launched in 2010 by the Norwegian FFI with the support of the Norwegian Space Centre and ESA \cite{eriksen2010tracking}. AISSat-1 (the first satellite of the AISSat constellation) was a 20 cm cube nano-satellite in sun-synchronous near-polar orbit at 635 km altitude and was conceived for demonstrating space-based AIS reception at high latitudes. The second demonstrator, called NorAIS-1, was deployed as a payload of the European Columbus module of the International Space Station and, therefore, enable maritime surveillance at low- to mid-latitudes. 
However,  Sat-AIS systems, such as NorAIS-1 and AISSat-1, provide poor detection performance in areas with high traffic density, such as central Europe, the Gulf of Mexico and South-East Asia \cite{eriksen2010tracking}. Indeed, the AIS protocol uses a time division multiple access (TDMA) scheme to avoid simultaneous transmission within a certain area, which is referred to as \emph{organized area} and typically covers a region up to 50 nautical miles~\cite{hoye2004space,eriksen2018metrics}. However, transmission collision avoidance is ensured only within the same organized area as ships in different organized areas transmit independently of each other. A Sat-AIS system will see different organized areas due to the large FOV and, therefore, in ocean regions with sufficiently high traffic density it may happen that different transmissions are received simultaneously by the satellite, resulting in message losses. It is worth noting that the larger the FOV, the higher the probability of message collisions. 
%A possible solution based on clever antenna design has been proposed and demonstrated in \cite{hoye2004space}. 
Additionally, space-based AIS systems experience interference from terrestrial VHF communications and suffer from a much lower signal strength compared to coastal AIS due to the much larger range. All such degradation factors contribute to message collisions and their combination makes Sat-AIS correctly receive only a fraction of all messages broadcast within the FOV. These detrimental effects could be alleviated by the use of multiple channels and clever antenna design, e.g., directional antennas, at cost of a reduced coverage~\cite{hoye2004space}.

\begin{table}
\footnotesize
\caption{List of current/future spaceborne AIS missions.}
\label{Table:Sat-AIS_missions}
\centering
\resizebox{.95\columnwidth}{!}{
\begin{tabular}{l l c c c c c}
\toprule
\textbf{Mission}	& \textbf{\mulrows{Organization \\(Country)}}	& \textbf{\mulrows{Launch \\ year}}  & \textbf{Altitude} $[$km$]$ & \textbf{Inclination}   & \textbf{\mulrows{Number of \\ satellites}} \\
\midrule
\midrule
AprizeSat	& Aprize Satellite Inc. (USA) & 2009	   & $686$  & $98.13^\circ$ 	& 2\\
\midrule
exactView	& exactEarth (Canada) 	& 2009		  & $500-817$ & $8^\circ, \sim 98^\circ$ & 11\\
\midrule
AISSat		& NSC (Norway) 			& 2010		  & $635$   & $97.71^\circ$ & 2\\
\midrule
VesselSat	& LuxSpace (Luxembourg)	& 2011		   & $865$ 	& $20^\circ$ 	& 2\\
\midrule
AAUSat-3		& Aalborg University (Denmark) & 2013   & $781$ & $98.55^\circ$ & 1\\
\midrule
AISat-1		& DLR (Germany) 			& 2014		  & $650$ 	& $98.3^\circ$ & 1\\
\midrule
Lemur		& Spire Global Inc. (USA) 			& 2014		  & $500$ 	& $37^\circ-85^\circ$ & 90 +\\
\midrule
NorAIS-2		& FFI (Norway) 			& 2015 		  & $408$ 	& $51.6^\circ$ & 1\\
\midrule
NorSat		& NSC (Norway)			& 2017		  & $600$ & $98^\circ$& 2\\
\midrule
Iridium NEXT & \mulrows{exactEarth (Canada) - \\ L3Harris (USA)} & 2017 & $780$ & $86.4^\circ$ & 66\\
\midrule
KOMPSAT-6	& KORI (South Korea) & 2020 & $505$ & $97.42^\circ$ & 1\\
\midrule
\textsuperscript{3}Cat-4	& UPC (Spain) & 2020 		  & $400$ 	& $51.6^\circ$ & 1\\
\bottomrule
\end{tabular}
}
\end{table}

Moreover, most of Sat-AIS systems are conceived as secondary payloads of EO satellites, which are typically launched in sun-synchronous orbits. Such a peculiarity leads to significant time gaps (up to 9 hours) close to the Equator, whereas shorter time gaps are experienced at higher latitudes \cite{eriksen2018metrics}. Nonetheless, the launch of different AIS satellites in constellation formation could reduce the revisit time significantly~\cite{disimone2017sea,disimone2017gnss}.
Since the launch of the first successful demonstrators, there has been a growing interest in Sat-AIS systems, and several dedicated missions have become operational.  
A partial list includes the AISSat-2 (the second satellite of the AISSat constellation) and NorAIS-2 (which replaces NorAIS-1) provided by FFI, the commercial exactView micro-sat constellation by exactEarth, the Iridium NEXT constellation by exactView and L3Harris, the two-sat VesselSat constellation operated by LuxSpace, the American two-sat AprizeSat constellation, Spire Global's Lemur constellation of more than 90 nanosatellites, the AISat-1 launched in 2011 by the German Aerospace Centre (DLR). In addition to dedicated instruments, AIS transponders are carried by different EO satellites, such as the U.K. NovaSAR-1 launched in late 2018 \cite{whittaker2011affordable} and the planned \textsuperscript{3}Cat-4 operated by the Polytechnic University of Catalonia \cite{munoz20183cat4}. A list of current and future spaceborne Sat-AIS missions with the respective technical details can be found in Table~\ref{Table:Sat-AIS_missions}.

Useful metrics for the evaluation of service quality offered by Sat-AIS are presented in \cite{eriksen2018metrics}, and include the overall number of collected messages and ships detected on a global scale and per time unit. Examples of global maps of detected messages and ship density are shown in \cite{helleren2012aissat} and \cite{buursink2012vesselsat}. Methods for assessing detection probability of Sat-AIS systems are also discussed in \cite{skauen2016quantifying} and \cite{li2018statistical}.
Despite the advantages of Sat-AIS with respect to coastal AIS in terms of coverage, reliable and seamless MS cannot rely on AIS protocols alone. Indeed, MS using AIS is limited to cooperative ships and large ships mounting properly-working AIS facilities onboard. Moreover, AIS
lacks authentication and encryption, and is hence
vulnerable to hacking or spoofing. Nonetheless, AIS relies on GPS which itself can be spoofed as reported in~\cite{GPSspoofing}.
Therefore, the monitoring of non-reporting ships is a crucial limitation of AIS and calls for additional information sources provided by remote sensing technologies.

\section{SAR}
\label{sec:overview_SAR}

\begin{figure}[!t]
\centering
\includegraphics[width=0.6\columnwidth]{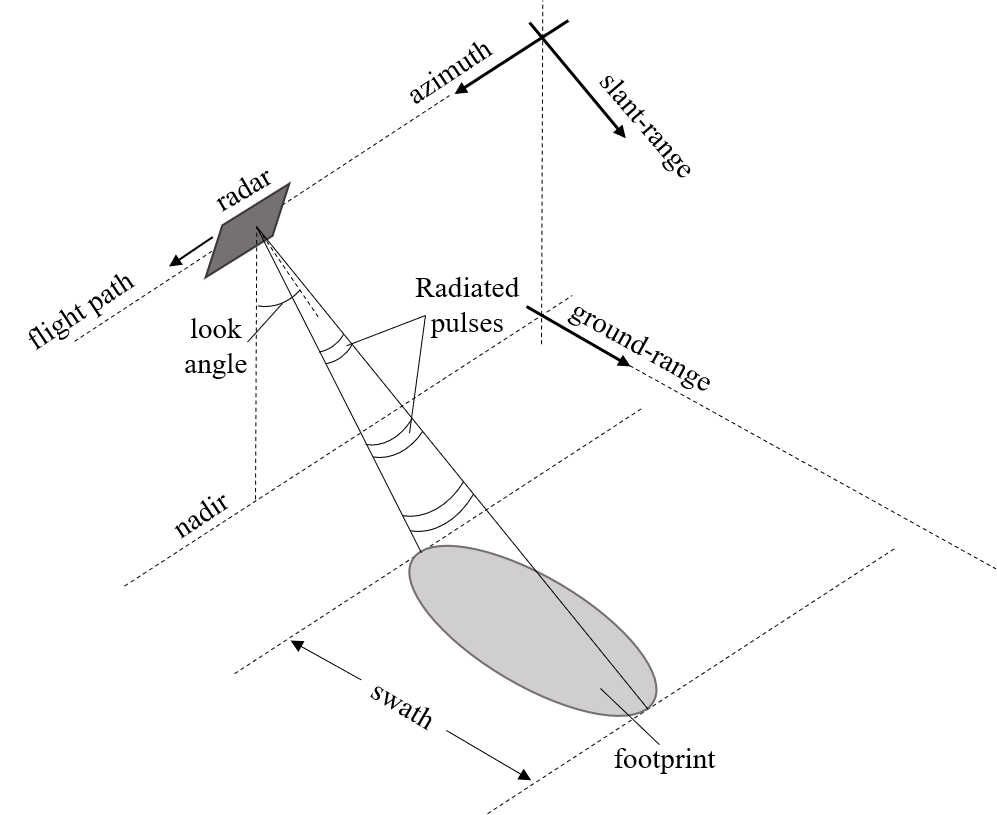}
\caption{Pictorial representation of a typical SAR acquisition geometry. The SAR sensor sends EM pulses at microwaves towards the Earth's surface and collects the backscattered signal. The side-looking geometry allows for unambiguous geolocation of the reflected echoes. The chirp signaling and the SAR processing improve the range and azimuth resolution, respectively, with respect to real aperture radars.}
\label{fig:SAR_geometry}
\end{figure}

SAR is a high-resolution imaging system typically employed on board satellites or aircraft. It is an active microwave remote sensing technology, i.e., it is equipped with a transmitter that sends electromagnetic (EM) microwave signals towards the Earth' surface. A typical SAR acquisition geometry is represented in Fig.~\ref{fig:SAR_geometry}. The presence of a dedicated transmitter enables a monitoring capability that does not depend on sunlight illumination and 
allows for effective sensing regardless of weather, as microwaves undergo a much weaker scattering and absorption from the atmosphere.
SAR systems typically operate in a monostatic geometry, where the receiving antenna is co-located with the transmitting one. In this configuration, only the energy reflected in the {\em backscattering} direction is collected by the system.
The received signal then undergoes a proper processing chain to form a 2-dimensional (2D) image of the scene reflectivity in slant-range/azimuth coordinates~\cite{franceschetti1999synthetic}. The features of the obtained image largely depend on both surface parameters, e.g., material composition, small- and large-scale roughness, and sensor parameters, e.g., viewing angle, frequency, polarization and spatial resolution.
Ship detection for MS is a mature application of SAR imagery and numerous automatic detectors have been developed so far~\cite{crisp2004state}. 
A detection is declared via constant false alarm rate (CFAR) processing, that is, from a thresholded departure above the ambient (noise plus clutter) as estimated from within a window surrounding the image sample under investigation~\cite{crisp2004state}.
More refined CFAR approaches adopt a parametric distribution model for the sea clutter to account for inhomogeneous sea state conditions \cite{crisp2004state}. 
However, despite good efficiency and reliability, which make CFAR techniques well-suited to near-real time MS, CFAR performance largely depends upon accuracy of the adopted sea clutter distribution model. 
%Apart from the classical Rayleigh distribution which is accurate for modeling amplitude of sea clutter in low spatial resolution images, other probability density functions, including the $\beta$, lognormal, Weibull, K and generalized-K distributions, have been successfully employed in CFAR applied to high-resolution imagery \cite{jakeman1987generalized}, \cite{ward2006sea}, \cite{jakeman1976model}. 
Indeed, CFAR performance may be significantly affected by the granular salt-and-pepper noise, referred to as \emph{speckle} noise, typical of SAR systems. 
%Indeed, the signals reflected from the elementary scatterers within the resolution cell are summed up coherently by the SAR receiver. This coherent combination leads to salt-and-pepper noise, referred to as \emph{speckle}, appearing in SAR imagery. 
Speckle noise is a detrimental effect, which significantly impacts both automatic processing by computer programs and interpretation by even SAR-expert human users. 
%Accordingly, a proper despeckling procedure is often applied as a pre-processing step to improve information retrieval. This is of key relevance in MSA as speckle noise may cause larger false alarm rate in CFAR techniques. 
In the recent past, many approaches have been developed for despeckling purposes. The most reliable methods are those based on non-local means, which exploit the redundancy inherent to SAR imagery, achieving speckle reduction without sacrifice of detail~\cite{parrilli2011nonlocal,deledalle2014exploiting,disimone2016scattering}. 
%\cite{deledalle2014exploiting}, \cite{parrilli2011nonlocal}, 
An additional feature offered by SAR systems is polarimetry. Polarimetric SAR (polSAR) is drawing increasing attention, especially for MS~\cite{liu2007polsar}. Indeed, fully-polarimetric SAR can provide more target scattering information than single- and dual-polarization SAR~\cite{ulaby1990radar,buono2016analysis}. 
However, despite the richer target information available with polSAR, the selection of polarimetric features suited to MS and detection of ships under high sea state conditions are still open problems. In addition, compared with optical sensors, polSAR images typically exhibit lower resolution. As a consequence, even advanced machine learning architectures %such as region-based convolutional neural networks (R-CNNs) \cite{ren2017rcnn}, 
that rely on target size may fail when using polSAR imagery. 
%In this regard, the exploitation of contextual semantic information has been demonstrated to improve detection performance \cite{jin2020patch}.

For MS purposes, both single-channel SAR and polSAR suffer from azimuth ambiguities, which arise due to the finite pulse repetition frequency combined with the strong returns from the sidelobes of the receiving antenna. Indeed, from a signal processing viewpoint, they can be regarded as an aliasing effect of the Doppler phase history. Their impact on target detection applications
can be severe, as multiple ``ghost'' replicas of a brilliant target (ships, oil platforms, bridges) may be detected in low backscatter areas (e.g., oceans in low wind speed regimes)~\cite{john1991synthetic}. There is already an extensive literature aimed at suppressing ghosts both in the time and frequency domains~\cite{livingstone2004focusing,moreira1993suppressing,guarnieri2005adaptive,dimartino2014filtering}. Classical approaches are based on the deconvolution of azimuth ambiguities with an ideal impulse response function \cite{moreira1993suppressing}. Other approaches recursively search for target replicas within the image by taking advantage of the dependence of target azimuth shifts upon SAR system parameters \cite{chen2013mitigation}. However, both deconvolution and search-based approaches are suited to point targets, such as small ships or ships in low-resolution images, whereas for extended targets selective filters must be applied \cite{li1983ambiguities,guarnieri2005adaptive,DiMartinoIRR14_7}. Such algorithms act as band-pass filters by selecting the portion of azimuth spectrum that is less affected by ambiguities. This allows good suppression capabilities also in the presence of extended targets, but naturally at a cost of reduced spatial resolution.

\begin{table}
\footnotesize
\caption{List of current/future spaceborne SAR missions (* Single (S), dual (D), tri (T), quad (Q)).}
\label{Table:SAR_missions}
\centering
\resizebox{.95\columnwidth}{!}{
\begin{tabular}{l l c c c c c}
\toprule
\textbf{Mission} 	  & \textbf{\mulrows{Country \\ (Organization)}} & \textbf{\mulrows{Launch \\ year}}  & \textbf{Band}   & \textbf{\mulrows{Incidence \\ angle}} 		& \textbf{Polarization$^{*}$}		& \mulrows{\textbf{Spatial} \\ \textbf{resolution} $[$m$]$}\\
\midrule
\midrule
TerraSAR-X 	  & Germany (DLR)		& 2007 		 & X & $15^\circ - 60^\circ$ & S, D, Q 	& $\geq$ 1\\
\midrule
COSMO-SkyMed   & Italy (ASI) 			& 2007 		 & X & $20^\circ - 59^\circ$ & S, D 		& $\geq$ 1\\
\midrule
RADARSAT-2 	  & Canada (CSA)			& 2007 		 & C & $20^\circ - 49^\circ$ & S, D, Q 	&  $\geq$ 3\\
\midrule
Sentinel-1 	  & Europe (ESA)			& 2014 		 & C & $29^\circ - 46^\circ$ & S, D 		& $\geq$ 5\\
\midrule
Gaofen-3		  & China (CNSA)			& 2016 		 & C & $10^\circ - 60^\circ$ & S, D, Q & $\geq$ 1\\
\midrule
SAOCOM		  & \mulrows{Argentina (CONAE) - \\ Italy (ASI)} & 2018 	 & L & $20^\circ - 50^\circ$ & S, D, Q & $\geq$ 10\\
\midrule
NovaSAR-1	  & UK (UKSA)			& 2018 		 & S & $15^\circ - 73^\circ$ & S, D, T 	& $\geq$ 6 \\
\midrule
COSMO-SkyMed 2 & Italy (ASI) 			& 2019 		 & X & $20^\circ - 59^\circ$ & S, D, Q 	& $\geq$ 0.35\\
\midrule
NISAR		  & \mulrows{USA (NASA) - \\ India (ISRO)} & 2022 		 & L, S & $33^\circ - 47^\circ$ & S, D, Q 	& $\geq$ 3\\
\midrule
Biomass		  & Europe (ESA)			& 2022 		 & P & $29^\circ - 46^\circ$ & Q			& $\geq$ 50 \\				
\bottomrule 
\end{tabular}
}
\vspace{1mm}
\end{table}

\begin{figure}[!t]
\centering
\includegraphics[width=0.5\columnwidth]{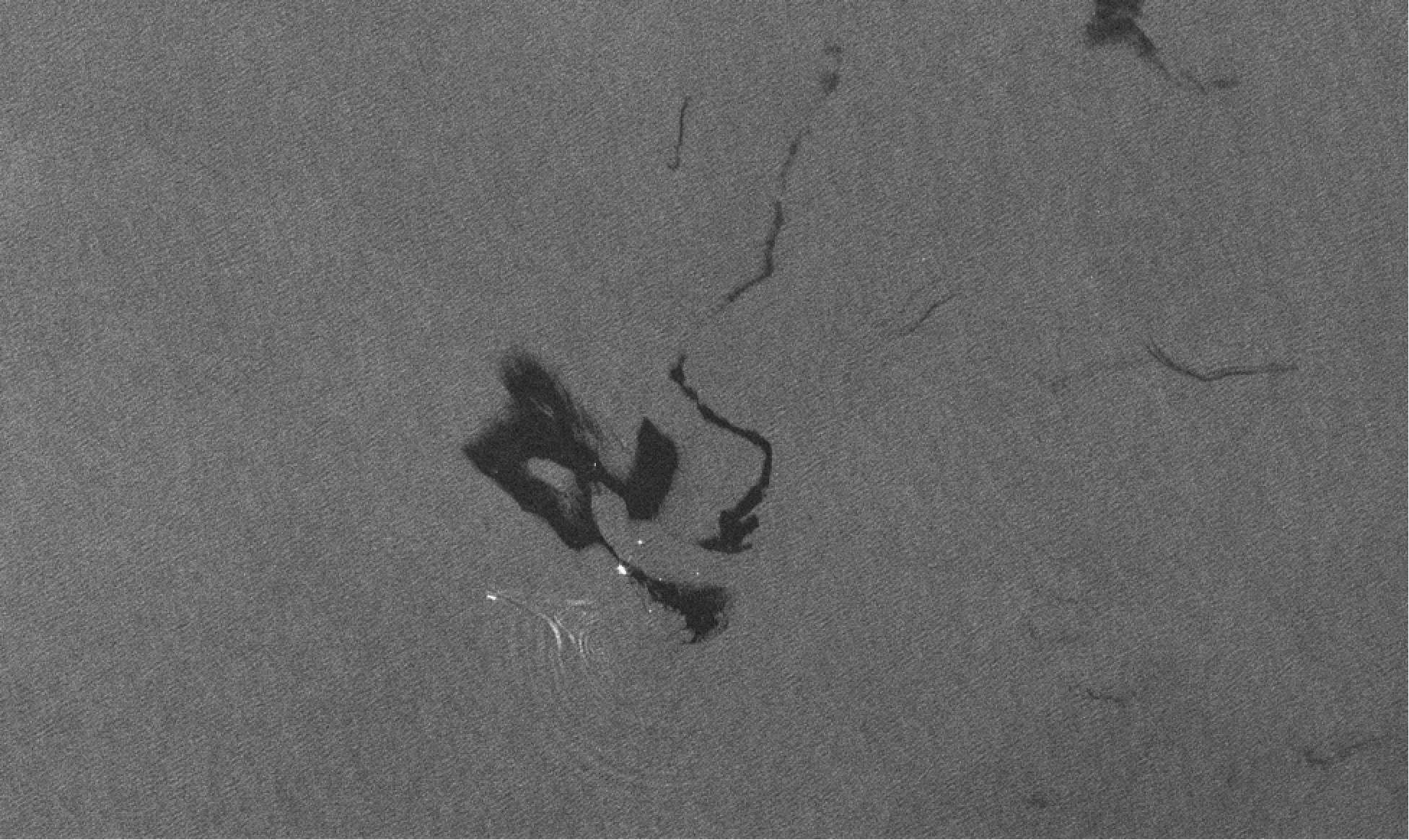}
\caption{Oil spill detection on a SAR image acquired by S-1 constellation in IW mode (10m resolution).}
\label{fig:oil_spill_det}
\end{figure}

Among past and current SAR missions, the Sentinel-1 (S-1) mission is particularly attractive for MS, thanks to free availability of data, low revisit time and polar orbit that ensures high-latitude coverage. The operating modes of S-1 are well suited to maritime activity monitoring: the Extra Wide (EW) swath mode is the typical acquisition mode over oceans and polar zones as images are collected over wide areas using five sub-swaths. EW images typically cover a 400 km swath at 20 m by 40 m spatial resolution with both dual-pol and single-pol options. It is then the preferable acquisition mode if the monitoring of large regions is required. Conversely, the Interferometric Wide (IW) mode is the main acquisition mode over land or restricted sea areas and allows for the detection of small targets thanks to the high spatial resolution. An example of a SAR image acquired in IW mode can be seen in Figure~\ref{fig:oil_spill_det}.
S-1 adopts the terrain observation with progressive scans SAR (TOPSAR) technique where the steering of the antenna beam in the range direction is accompanied by an additional electronic beam steering in the azimuth direction from backward to forward. 
The range steering allows TOPSAR to achieve the same coverage and resolution of ScanSAR, whereas the azimuth steering allows for a more homogeneous signal-to-noise ratio (SNR) over the whole coverage area.

\begin{figure}[!t]
\centering
\includegraphics[width=0.5\columnwidth]{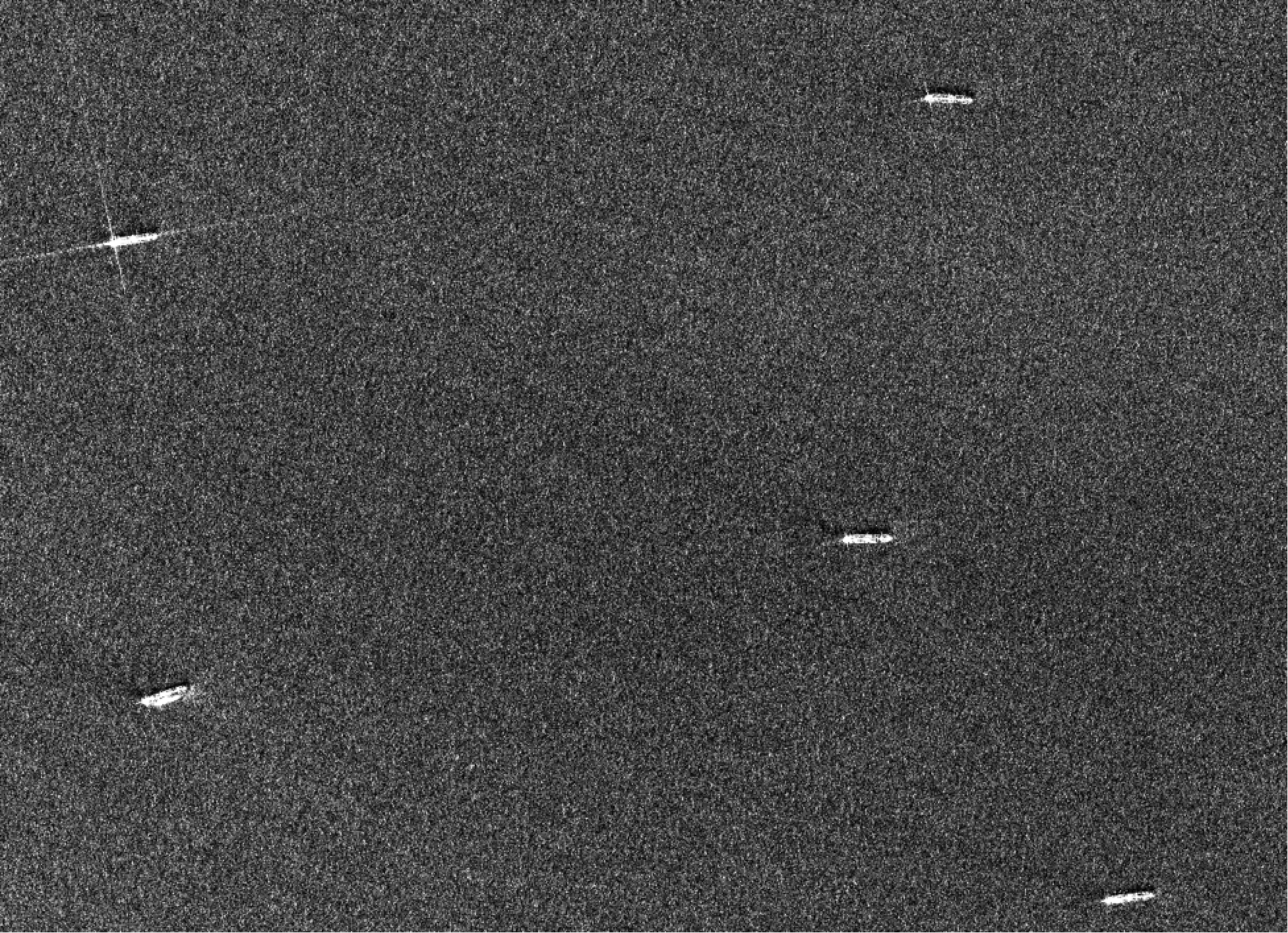}
\caption{Vessels detected using a SAR image acquired by COSMO-SkyMed constellation in Stripmap HIMAGE mode (5m resolution).}
\label{fig:SAR_ship_det}
\end{figure}

\begin{figure}[!t]
\centering
\includegraphics[width=0.5\columnwidth]{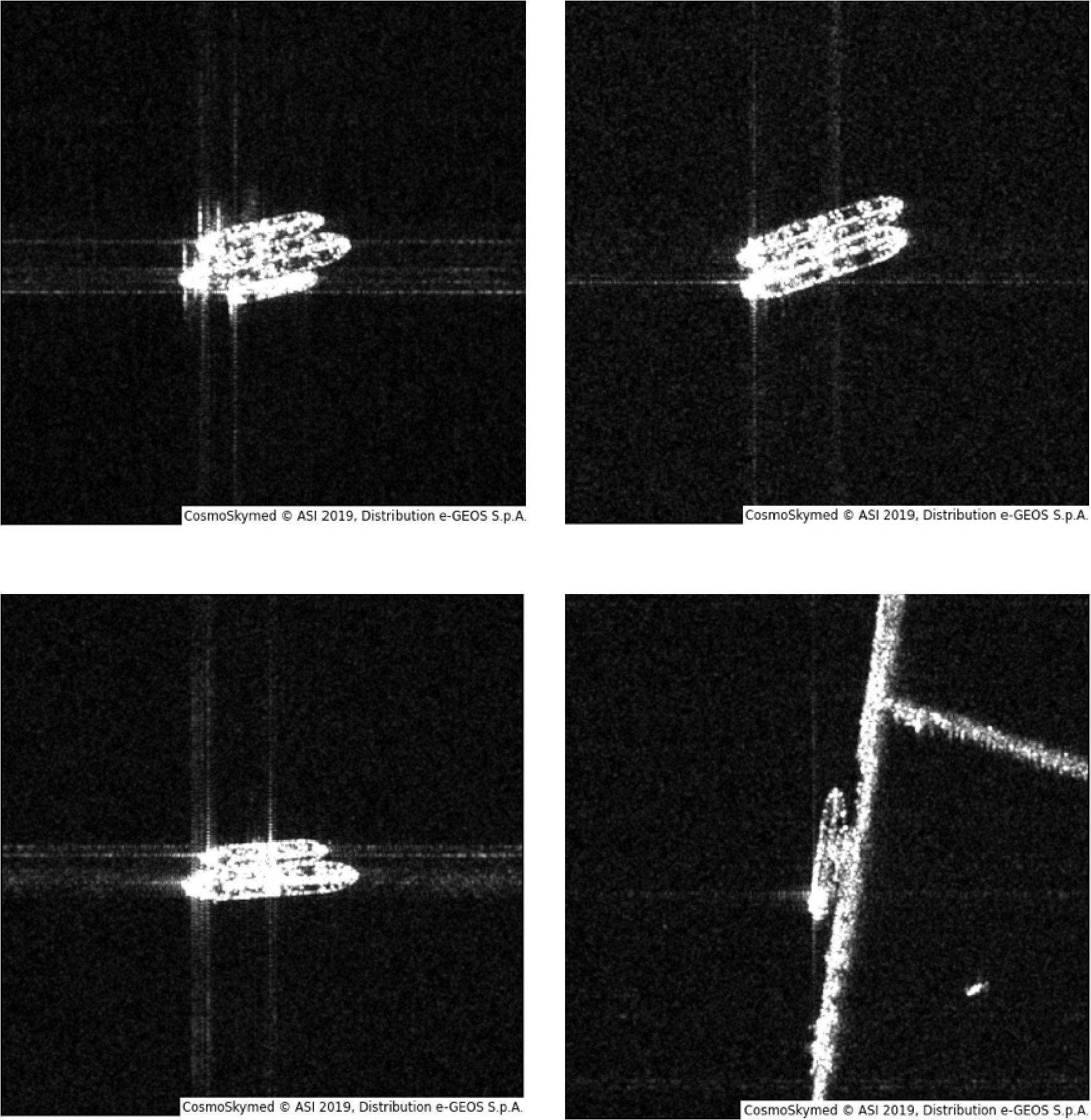}
\caption{Example of vessels without AIS and anomalous behaviour (transhipment) as detected using a set of SAR images acquired by COSMO-SkyMed constellation in HI mode (5m resolution).}
\label{fig:SAR_image_cosmo}
\end{figure}

COSMO-SkyMed is an EO satellite space-based radar system funded by the Italian Ministry of Research and Ministry of Defence and conducted by the Italian Space Agency (ASI), intended for both military and civilian uses. The constellation includes four first-generation satellites and two second-generation satellites, already in orbit, following sun-synchronous polar orbits and equipped with SAR systems with global coverage of the planet. Observations of an area of interest can be repeated several times a day in all-weather conditions. 
The COSMO-SkyMed satellite main payload is an X-band, multi-resolution and  multi-polarization imaging radar that, thanks to the implementation of different acquisition modes, can guarantee both wide area coverage at medium/high resolution, i.e., Stripmap HIMAGE (HI) / SCANSAR modes, and narrow area coverage at very high resolution, i.e., Spotlight-2 mode, thus providing  contributions at different scales to MS.
%\textcolor{red}{The typical acquisition mode for MSA purposes of COSMO-SkyMed is the Stripmap HIMAGE (HI) mode. SAR images acquired in this mode typically cover a 40 km swath at 5~m by 5~m spatial resolution}. 
Example of SAR images acquired in HI mode by the COSMO-SkyMed constellation are shown in Figures~\ref{fig:SAR_ship_det} and~\ref{fig:SAR_image_cosmo}. 
A list of current and future spaceborne SAR missions with the respective technical details can be found in Table~\ref{Table:SAR_missions}.
%The European Maritime Safety Agency (EMSA) is providing to European member states the CleanSeaNet service, that is the Near Real Time European satellite based oil spill monitoring and vessel detection service, set up and operated by the EMSA since April 2007. The service is able to provide warnings on possible oil spills, pollution alerts and related information to the operational maritime administrations within 30 minutes of the appearance of an issue, after satellite acquisition to allow an effective use of the data for maritime surveillance. 
%The downlink of satellite acquired data to the ground may not always follow the strategy of the nearest ground station but also use other satellites (e.g. geostationary telecommunication satellites) as space relay node for forwarding the collected information to the target ground station, without expecting to be in its field of view. 

\section{MSP Optical sensors}
\label{sec:MSP_optical_sensors}
\begin{figure}[!t]
\centering
%\begin{postscript}
%\psfrag{unit}[c][c][.80]{\raisebox{0mm}{\hspace{0mm}Wavelength ($\mu$m)}}
\includegraphics[width=0.65\columnwidth]{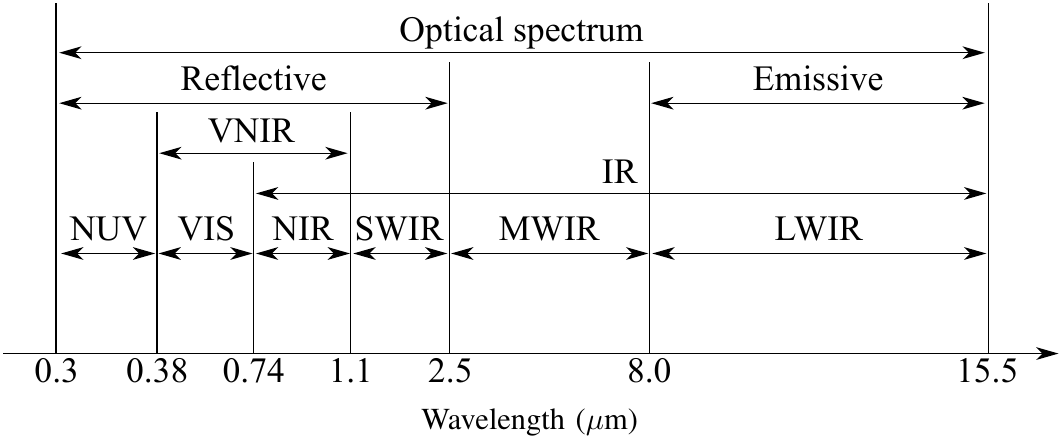}
%\end{postscript}
\caption{Optical spectrum with different bands indicated. The following acronyms are used: near-ultraviolet (NUV), visible (VIS), near-infrared (NIR), shortwave infrared (SWIR), midwave infrared (MWIR), longwave infrared (LWIR), VIS and NIR (VNIR) and infrared (IR).}
\label{fig:optical_spectrum}
\end{figure}

\begin{figure}[!t]
\centering
\includegraphics[width=0.5\columnwidth]{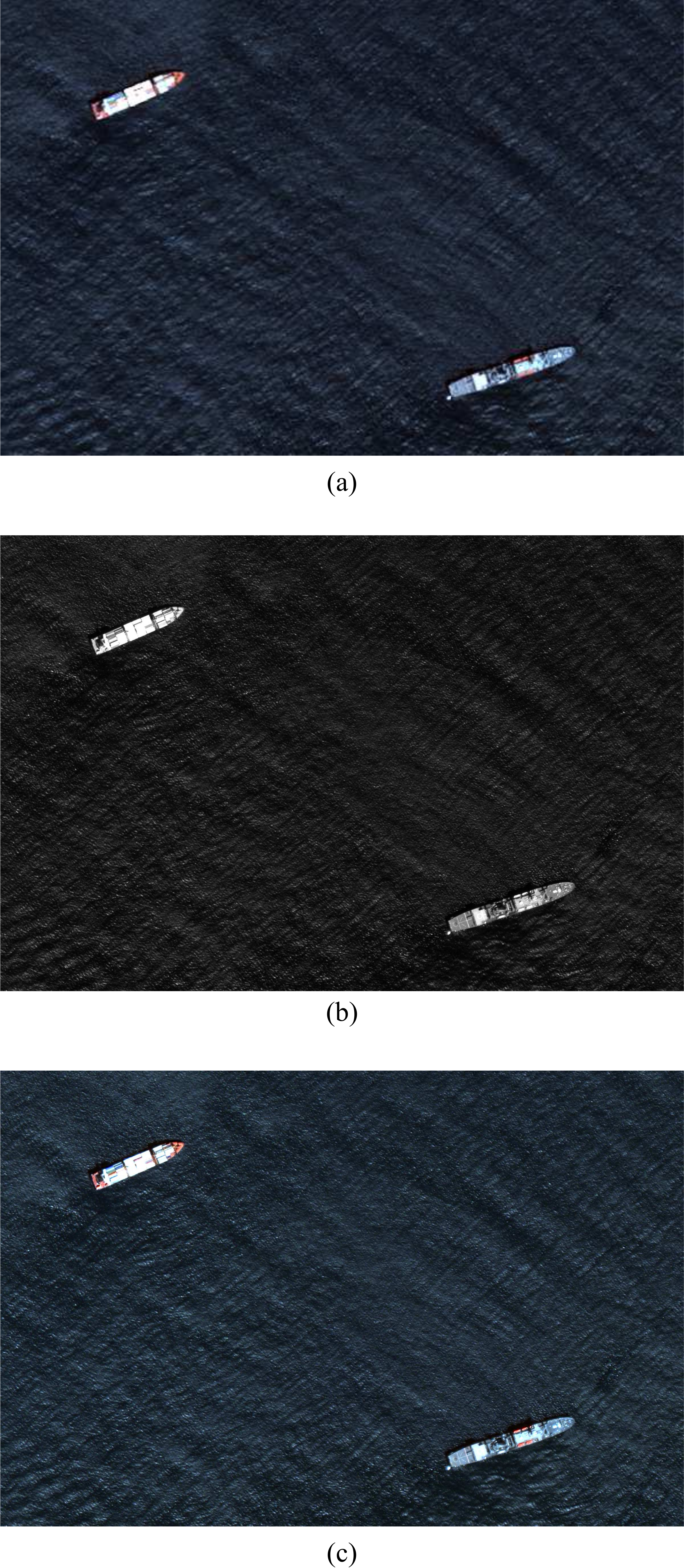}
\caption{(a) Detected vessels in a high-resolution MSP image (RGB) acquired by WorldView-2 satellite (2~m resolution). (b) Same detected vessels in a high-resolution PAN image (0.5~m resolution). (c) Pansharpened (RGB + PAN) image showing the two detected vessels.}
\label{fig:MSP_examples}
\end{figure}

While active microwave remote sensing systems, such as SAR, offer a key support for gathering rather coarse information at multiple frequencies, polarizations, and viewing angles, MSP and HSP optical sensors can be exploited fruitfully to infer more accurate details in the spatial and spectral domains. 
%MS and HS sensors cover the entire optical region which is bounded between microwaves and X-rays and refers to wavelengths ranging from $0.3$ $\mu\text{m}$ to $15$ $\mu\text{m}$ and includes near-ultraviolet ($0.3$ $\mu\text{m to } 0.38$ $\mu\text{m}$), visible (VIS) ($0.38$ $\mu\text{m to } 0.74$ $\mu\text{m}$), VIS near infrared (VNIR) ($0.74$ $\mu\text{m to } 1.1$ $\mu\text{m}$), shortwave infrared (SWIR) ($1.1$ $\mu\text{m to } 2.5$ $\mu\text{m}$), midwave infrared (MWIR) ($2.5$ $\mu\text{m to } 8$ $\mu\text{m}$) and longwave infrared (LWIR) ($8$ $\mu\text{m to } 15$ $\mu\text{m}$), also referred to as thermal infrared, \cite{landgrebe2005signal}, see Fig.~\ref{fig:optical_spectrum}.
%Remote sensing technologies operating in these bands are passive systems that detect and measure the sunlight radiation reflected or thermal emission by objects.
% 
%VNIR and SWIR parts of the optical spectrum have experienced a particular interest in the military context as they allow for the discrimination of 
%camouflaged objects \cite{shimoni2019hypersectral}.
MSP and HSP sensors cover the entire optical region -- between microwaves and X-rays -- and refers to wavelengths ranging from 0.3 $\mu$m to 15 $\mu$m\footnote{This includes near-ultraviolet (NUV) (0.3 $\mu$m to 0.38 $\mu$m), visible (VIS) (0.38 $\mu$m to 0.74 $\mu$m), near-infrared (NIR) (0.74
$\mu$m to 1.1 $\mu$m), shortwave infrared (SWIR) (1.1 $\mu$m to 2.5 $\mu$m), midwave infrared (MWIR) (2.5 $\mu$m to 8 $\mu$m) and longwave infrared (LWIR) (8 $\mu$m to 15 $\mu$m), also referred to as thermal infrared~\cite{landgrebe2005signal}. VIS and NIR (VNIR), and SWIR parts of the optical spectrum have evoked particular interest from the military, as they allow for the discrimination of camouflaged objects~\cite{shimoni2019hypersectral}.}\cite{landgrebe2005signal} as shown in Fig.~\ref{fig:optical_spectrum}.
Most space-based MSP instruments operated by national or international space agencies, e.g., Landsat, Sentinel-2 (S-2), SPOT, provide images at spatial resolutions similar to SAR systems (decametric to metric). However, an increasing number of commercial missions can currently offer optical imagery at resolutions down to few tens of centimeters, e.g., WorldView-3, ALOS-3, SPOT-7, RapidEye.
A list of current and future spaceborne MSP missions with the respective technical details can be found in Table~\ref{Table:MSP_missions}.
Space-based MSP sensors typically acquire information in few ($< 10$) bands \cite{yokoya2017hyperspectral}. Among them, MSP data comprise the red ($0.64 - 0.67$ $\mu$m), green ($0.53 - 0.59$ $\mu$m) and blue ($0.45 - 0.51$ $\mu$m) bands, which are combined in RGB channels to form optical images.
Additional bands typically include: a panchromatic channel measuring a wider portion of the visible spectrum and, therefore, exhibiting higher spatial resolution than RGB channels; an ultra-blue channel ($0.40 - 0.45$ $\mu$m), also referred to as coastal/aerosol band, which is mainly exploited for analyzing shallow water composition and essaying atmospheric particles, like dust and smoke; one or multiple channels measuring VNIR, which provide useful information about plant health and is, therefore, mainly used for studying vegetated areas, e.g., through normalized difference vegetation index; and some bands covering portions of SWIR suited to such geological applications rock- and soil-typing. MWIR bands are rarely exploited. 
%(e.g., in MODIS) as sunlight radiation has fallen to a negligible level and thermal radiation from objects upon the Earth's surface has not yet achieved an appreciable level \cite{iqbal2012introduction}.
\begin{table}
\footnotesize
\caption{List of current/future spaceborne MSP missions.}
\label{Table:MSP_missions}
\centering
\resizebox{.95\columnwidth}{!}{
\begin{tabular}{l l c c c c c}
\toprule
\textbf{Mission} & \textbf{\mulrows{Organization \\ (Country)}} & \textbf{\mulrows{Launch \\ year}} & \mulrows{\textbf{Spectral} \\ \textbf{range} $[\mu$m$]$} & \textbf{Bands} & \mulrows{\textbf{Radiometric} \\ \textbf{resolution} $[$bit$]$} & \mulrows{\textbf{Spatial resolution} \\ \textbf{(MS/PAN)} $[$m$]$}\\
\midrule
\midrule
RapidEye 	& \mulrows{Planet Labs \\ (Germany)} 	& 2008 		 & $0.44 - 0.85$  & 5 		& 12 					& 6.5\\
\midrule
ResourceSat-2& ISRO (India) 			& 2011		 & $0.52 - 1.70$ & 4 + PAN	& 10 					& $\geq$ 5.8 (PAN + MS) \\
\midrule
KOMPSAT-3	& \mulrows{KORI \\ (South Korea)} 	& 2012     	 & $0.45 - 0.9$  & 4 + PAN 	& 14 				& 2.8 (MS) / 0.7 (PAN)\\
\midrule
Landsat 8 	& NASA (USA)			  	& 2013 	   	 & $0.433 - 2.3$ & 8 + PAN  	& 12				  	& 30 (MS) / 15 (PAN)\\
\midrule
SPOT-7		& \mulrows{EADS Astrium  \\ (France)}  	& 2014 	   	 & $0.45 - 0.89$ & 4 + PAN  	& 12 				  	& $\geq$ 6 (MS) / 1.5 (PAN)\\
\midrule
WorldView-3 	& \mulrows{DigitalGlobe \\ (USA)}		& 2014 		 & $0.4 - 2.365$ & 28 + PAN  & \mulrows{11 (PAN + MS) \\ 14 (SWIR)} & $\geq$ 1.24 (MS) / 0.31 (PAN)\\
\midrule
Gaofen-2 	&  CNSA (China) 			& 2014 		 & $0.45 - 0.89$	 & 4 + PAN 	& 14  				& 3.2 (MS) / 0.8 (PAN)\\
\midrule
Sentinel-2 	& ESA (Europe) 		  	& 2015 	   	 & $0.4 - 2.4$ 	 & 13 	  	& 12  				& $\geq$ 10 (MS)\\
\midrule
%Terra-ASTER 	& USA (NASA) - Japan (J-Space Systems)	&  &  &  & & \\
%\midrule
DMC-3 		& \mulrows{DMC International \\ Imaging (UK)} & 2015 & $0.44 - 0.91$ & 4 + PAN & 10  & 4 (MS) / 1 (PAN)\\
\midrule
ALOS-3 		&  JAXA (JAPAN)			& 2021 		 & $0.52 - 0.77$ & 6 + PAN   & 11  				& 3.2 (MS) / 0.8 (PAN)\\
\bottomrule 
\end{tabular}
}
\end{table} 
Finally, LWIR bands are exploited in some missions (e.g., Landsat) to measure Earth' surface temperature.
Despite the large number of bands available in MSP data, MS applications, e.g., ship detection/tracking, target identification, oil spill detection, make typically use of either RGB~\cite{wu2009performance,daniel2013automatic,heiselberg2016direct,park2018multi} or panchromatic (PAN)~\cite{corbane2008using,proia2009characterization,zhu2010novel,corbane2010complete,yang2013ship,shi2013ship,tang2014compressed,zou2016ship} images. 
Thanks to its higher spatial resolution, PAN data allow for detection of smaller targets compared to RGB channels, which, conversely, are more suited for target classification purposes, owing to the more refined spectral information. However, detection of small ship targets (on the order of meters) is still challenging and, therefore, finer spatial details are typically preferred over additional colour bands. This has led to much larger efforts by the scientific community in exploiting PAN or PAN + MSP~\cite{VivoneACDGLRW15_53} rather than only RGB for ship detection purposes \cite{kanjir2018vessel}. 
Fig.~\ref{fig:MSP_examples} shows detected vessels in a MSP image acquired by WorldView-2. More specifically, Fig.~\ref{fig:MSP_examples}-(a) depicts a very high-resolution RGB image of two vessels acquired in the Gulf of Naples, Italy, whereas Fig.~\ref{fig:MSP_examples}-(b) shows the corresponding PAN band. 
The spatial resolution of the RGB and PAN images is 2 m and 0.5 m, respectively. RGB and PAN bands are combined through pansharpening~\cite{VivoneACDGLRW15_53} in Fig.~\ref{fig:MSP_examples}-(c) to enhance the spatial resolution of the RGB image.
As opposed to VIS spectrum, LWIR bands do not depend on solar illumination as they rely on spontaneous thermal emission by objects. Therefore, they are more suited to night-time monitoring, even if a temperature contrast between target and background sufficient for detection purposes might be experienced during day. 
Unfortunately, LWIR bands offer a coarser spatial resolution with respect to VIS bands (e.g., 100 m against 30 m in Landsat 8 imagery) and exhibit a larger sensitivity to atmosphere moisture content.
%These drawbacks explain the rather limited literature focusing on the exploitation of thermal infrared data \cite{wang2017detecting}, \cite{yang2018ship}. 

An extensive review of the works published from 1978 to 2017 and focused on vessel detection and classification using optical sensors data for MS applications is presented in \cite{kanjir2018vessel}.
Approaches to vessel detection using optical satellite imagery can be grouped into several classes as illustrated in \cite{kanjir2018vessel}. However, most common techniques include threshold-based methods~\cite{corbane2008using,yang2013ship}, where candidate targets are selected by comparing image pixels with a properly evaluated (possibly local) threshold, and methods based on shape and texture features~\cite{zhu2010novel,corbane2010complete,shi2013ship}, where potential targets are extracted by taking advantage of the different geometrical and spectral characteristics of ships versus the surrounding sea.
The richer spectral information provided by optical sensors compared to SAR can be exploited fruitfully for target classification beyond simple size-based categorization towards more refined vessel type identification \cite{kanjir2018vessel}.  
More recently, the increasing amount of free and open optical data, e.g., S-2, Landsat 8, along with the advances in hardware (e.g., GPU computing), has boosted the application of machine learning paradigms, e.g., deep learning, convolutional neural networks and support-vector machines, to both detection and classification purposes for MS (see further details in the companion paper~\cite{SpaceAESM_2:J20}).
To summarize, a high level of spatial detail along with the easier interpretability and the absence of speckle noise make optical satellite imagery a very attractive solution for maritime surveillance purposes. 
However, compared to SAR, MSP imagery is affected by weather conditions, e.g., clouds, and ocean waves.

\section{HSP Optical Sensors}
\label{sec:HSP_optical_sensors}
As opposed to MSP instruments, HSP remote sensing is a relatively new technology able to detect and measure target radiance up to hundreds of bands over the visible/infrared (IR) portion of the EM spectrum.
HSP systems offer a very high spectral resolution (less than 1 nm), which allows near-laboratory quality measurements of radiation from far distances and a ``continuous'' mapping of the spectral signature of the sensed object compared with the discrete spectral sampling provided by MSP sensors. This peculiar feature enables the quantitative analysis and identification of any substance --- such as minerals, chlorophyll, chemical agents and gases by comparing its measured spectral profile with prepopulated libraries containing spectral signatures of different targets.
Accordingly, HSP data are typically exploited in applications such as natural hazard forecasting, atmosphere analysis, precision farming, water quality control, urban mapping and snow monitoring~\cite{bioucas2013hyperspectral}.
The ability to reveal molecular absorption and transmissivity bands with fine spectral details, unattainable by current MSP sensors, makes optical HSP sensors also referred to as imaging spectrometers, in analogy with chemical spectroscopy \cite{landgrebe2005signal}.

HSP systems offer a unique tool in the framework of MS, thanks to the high spectral resolution that adds a further dimension to space-based measurements. Indeed, HSP imagery extends surveillance and monitoring capabilities by providing key support for detecting difficult targets, i.e., objects that are smaller, more fleeting, and/or are immersed in more complex clutter, than those typically detectable with other remote sensing technologies. 
For instance, in maritime environments, typical low-observable targets detectable using HSP data include chlorophyll and sediment load in physical oceanographic applications, submarines and sea mines in military applications and schools of cetaceans in marine biology applications. In all these scenarios, the detection problem refers to low-contrast targets immersed in high-clutter background, whereas detection performance depends critically on the spectral properties of seawater.  Some tools, such as the portable profiling oceanographic instrument system (PorPOIS) by the Space and Naval Warfare Systems Center of San Diego, CA, US, were developed to provide a thorough characterization of the marine environment \cite{stein2001hyperspectral}.
Additionally, HSP systems allow for identifying contaminants, detecting underwater targets, enhancing accuracy in image classification and strengthening anti-camouflage capabilities \cite{stein2001hyperspectral}.
\begin{table}
\footnotesize
\caption{List of current spaceborne HSP missions}
\label{Table:HSP_missions}
\centering
\resizebox{.95\columnwidth}{!}{
\begin{tabular}{l l c c c c c}
\toprule
\textbf{Mission} 	& \textbf{\mulrows{Organization \\ (Country)}} & \textbf{\mulrows{Launch \\ year}} & \mulrows{\textbf{Spectral range} $[\mu$m$]$ \\ \textbf{(Number of bands)}}& \mulrows{\textbf{Spectral} \\ \textbf{resolution} $[$nm$]$} & \mulrows{\textbf{Radiometric} \\ \textbf{resolution} $[$bit$]$} & \mulrows{\textbf{Spatial} \\ \textbf{resolution} $[$m$]$} \\
\midrule
\midrule
Hyperion 	& NASA (USA) 			& 2000  			& $0.357-2.576$ ($220$) & $10$ & 12 	& 30\\
\midrule
PROBA-1		& ESA (Europe)			& 2001			& $0.4-0.105$ ($\leq 62$) & $1.25 - 11$  & 12 &  $\geq 17$\\
\midrule
AIRS 		& NASA (USA) 			& 2002 			& \mulrows{VNIR: $0.4-0.94$ (4) \\ IR: $3.74-15.4$ (2378) }& IR: $4$ & - 	& \mulrows{VNIR: 2300\\ IR: 13500 }\\
\midrule
IASI 		& \mulrows{CNES/EUMETSAT \\ (Europe)} & 2006  & $3.62-15.5$ (8461) & $2.8$ & 16 	& 12000\\
\midrule
Rising-2 	& Japan 					& 2014 			& \mulrows{VIS: $0.4-0.65$ (3) \\ VNIR: $0.65-1.05$ (401)} & VNIR: 1 & 10 & 4.6 \\
\midrule
PRISMA 		&  ASI (Italy)		    & 2019 			& \mulrows{VNIR: $0.4-1.01$  (66) \\ SWIR: $0.92-2.5$ (173) \\ PAN: $0.4-0.7$} & $\leq 14$  & 12 & \mulrows{VNIR: $30$  \\ SWIR: $30$ \\ PAN: $5$} \\
\midrule
\textsuperscript{3}Cat-5/B	& \mulrows{ESA (Europe) - \\ Spain (UPC)} & 2020 & \mulrows{VNIR: $0.4-1$ (45) \\ LWIR: $8-14$ (3)} & VNIR: 16 & - & \mulrows{VNIR: 75 \\ LWIR: 490}\\
\midrule
EnMap		& DLR (Germany) 			& 2020			& $0.42-0.245$ (228) & \mulrows{VNIR: $6.5$ \\ SWIR: $10$} & 14 & 30  \\  
\bottomrule 
\end{tabular}
}
\end{table}
The promising identification capabilities of HSP led to an extensive literature dealing with the detection problem using HSP data \cite{manolakis2003hyperspectral,manolakis2013detection,matteoli2014overview,nasrabadi2013hyperspectral,axelsson2016target}.
A comprehensive collection of recent applications of HSP data in security and defence environments is available in \cite{shimoni2019hypersectral}.
The high dimensionality of HSP data calls for advanced image processing and allied big data analysis, as well as high data storage capabilities and computational needs. Consequently, the rapid increase of interest in HSP imaging systems only recently is not surprising~\cite{landgrebe2002hyperspectral}. Nevertheless, the usage of high-dimensional HSP data is still challenging in surveillance applications, where time constraints are strict. Recently, algorithms aimed at relaxing processing resources via dimensionality-reduction have been developed \cite{harsanyi1994hyperspectral,plaza2005dimensionality,li2018discriminant} but are still inadequate for real-time classification \cite{axelsson2016target}.
Well-developed military applications of HSP data, including space surveillance, camouflage countermeasures and detection of weapons of mass destruction, were also identified by the North Atlantic Treaty Organization (NATO), along with the European Defence Agency and the U.S. Department of Defense \cite{shimoni2019hypersectral}.
Keeping in mind the unavoidable tradeoff between spatial and spectral resolution, optical remote sensing systems offer complementary features: MSP sensors acquire information at very high spatial resolution but with large spectral separation between bands larger compared to HSP systems that, conversely, provide fine spectral details at a limited (typically decametric) spatial resolution.
In the recent past, this has motivated significant effort aimed at taking advantage of both HSP and MSP imagery. Accordingly, a wide literature has been focusing on the enhancement of the spatial details of HSP imagery. The most common approaches include the fusion of either HSP and MSP data, also referred to as \emph{hypersharpening}, or HSP and panchromatic images, known as HSP pansharpening \cite{loncan2016hyperspectral}. Different methodologies, including multiresolution analysis, convolutional neural networks, spectral unmixing, Bayesian and matrix factorization, component substitution, are adopted to this end.  Deep comparative reviews are presented in~\cite{loncan2016hyperspectral,mookambiga2016comprehensive,yokoya2017hyperspectral}. 
The combination of complementary advantages offered by HSP and MSP systems through hypersharpening might enable material reconnaissance with unprecedented spatial resolution, thus offering an invaluable support to MS applications, e.g., object detection, scene interpretation and classification.
A list of current spaceborne HSP missions with the respective technical details can be found in Table~\ref{Table:HSP_missions}.

\section{GNSS-R}
\label{sec:overview_GNSSR}
Passive bistatic radar (PBR) systems have experienced growing interest in the scientific and defence communities only in the recent past, despite their history dating back to 1955, before the development of monostatic radars \cite{hanle1986survey,FarinaL14_8}. The main feature of PBR is passive mode, i.e., the capability to exploit external (i.e., non-cooperative) radiofrequency (RF) sources as signals of opportunity for remote sensing applications \cite{hanle1986survey}. Typical signal sources used by PBRs include radio and television services, e.g., DAB/DVB-T~\cite{conti2012high} and FM~\cite{dilallo2008design}, mobile communication systems, such as GSM~\cite{tan2005passive}, wireless networks, e.g., Wi-Fi~\cite{falcone2012potentialities}, WiMAX~\cite{higgins2016passive}.
Passivity brings numerous benefits with respect to other remote sensing technologies, such as SAR and MSP/HSP optical sensors. Indeed, the lack of a dedicated transmitter makes PBR inherently cost-effective, compact and frugal with energy since they comprise only the receiving unit. As a result, PBR might be deployed in network configurations consisting of many radar units located at different places and cooperating each other, i.e., as multi-static radar systems, at relatively low cost \cite{chernyak1998fundamentals}. Additionally, bi- and multi-static radars increase the level of safety and security in military contexts as the (high value) receiving stations can be located independently from a transmitter that may be peripheral to military's interest. Finally, the spatial separation transmitter-to-receiver and receiver-to-receiver allows more scattering information to be gathered through bi-/multi-static configurations with respect to monostatic geometry. All such aspects make PBR an intriguing technology in electronic warfare as it supports anti-stealth, anti-missiles, anti-jamming capabilities, and is suited in emission control contexts.
However, PBRs have a severe limitation: they do not allow for a proper design of the transmit waveform and power which, typically, are little suited for MS applications.
\begin{figure}[!t]
\centering
\includegraphics[width=0.5\columnwidth]{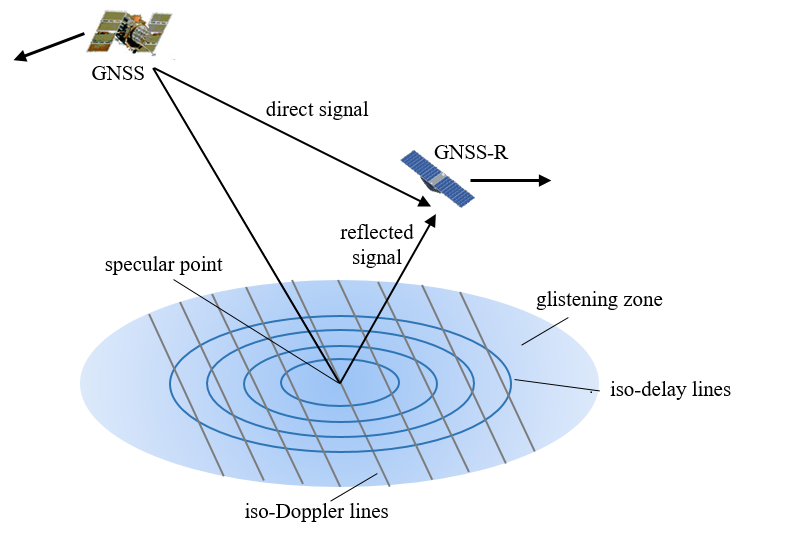}
\caption{Pictorial representation of GNSS-R basic principle. Navigation signals emitted from GNSS stations, e.g., GPS, GLONASS, Galileo, are reflected off the Earth's surface and acquired by the GNSS-R receiver. Due to the low SNR of Earth-reflected GNSS signals, standard GNSS-R receivers operate in a forward-scattering geometry, where most energy is collected from a region surrounding the specular-reflection point, i.e., the so-called glistening zone.}
\label{fig:GNSS_R}
\end{figure}

Additionally, terrestrial illuminators have poor penetration offshore and can be exploited as opportunistic RF sources only for limited MS applications, e.g., covert operations, harbor protection, shore and rivers traffic monitoring. Accordingly, for MS in open sea, it is mandatory to switch to satellite-based illuminators and receivers which can provide seamless coverage at global scale. In this regard, passive spaceborne systems, referred to as GNSS-R, exploiting GNSS signaling, represent an intriguing solution for space-based EO and MS. In addition to continuous and worldwide coverage capabilities, GNSS-based PBR can benefit, at least in principle, from the availability of more than 100 signal sources, including GPS, GLONASS, and the recently fully operating Galileo and BeiDou-2 GNSSs. 
Besides the original purposes of GNSS -- providing location and navigation
services to users located on or near the Earth's surface -- navigation signals emitted by GNSS stations have been opportunistically adopted in remote sensing and EO for MS. 
The working principle of GNSS-R is illustrated in Fig.~\ref{fig:GNSS_R}: navigation signals emitted from GNSS stations are collected 
by GNSS-R receivers after scattering from the Earth' surface. Due to the low signal strength, GNSS-R acquires the Earth-reflected 
signals coming from a region surrounding the specular reflection point, referred to as \emph{glistening zone}, where scattering is 
stronger \cite{zavorotny2014tutorial}. In order to further increase the strength of the signal reflected off the sea surface, the 
left-hand circular polarization (LHCP) channel is typically preferred to the much weaker right-hand circular polarization (RHCP) 
component \cite{zavorotny2014tutorial}. This is the typical configuration of spaceborne GNSS-R instruments and is referred to as 
standard GNSS-R hereafter. Direct signals coming from the transmitting antenna without reflection are measured for location 
and processing aims. By proper processing of the direct and reflected signals, geophysical parameters can be inferred~
\cite{zavorotny2014tutorial}.
Thanks to passivity, GNSS-R payloads are often hosted in small satellites, typically nanosats (mass up to 10 kg), or CubeSats, i.e., 
small satellites multiple of a standard satellite of size 10 cm $\times$ 10 cm $\times$ 11.35 cm, weighing less than 1.33 kg 
\cite{camps2019nanosatellites}. This has led to the development and launch of large constellations of small satellites equipped with 
a GNSS-R payload, such as the ongoing NASA CYGNSS eight-satellite mission for hurricane prediction \cite{ruf2013cygnss} and the 
planned commercial GNSS-R constellations by Spire Global, Inc. \cite{spire2019earth}. 
A list of current and future spaceborne GNSS-R missions with the respective technical details can be found in Table~\ref{Table:GNSS_R_missions}.
\begin{table}
\footnotesize
\caption{List of current/future spaceborne GNSS-R missions.}
\label{Table:GNSS_R_missions}
\centering
\resizebox{.95\columnwidth}{!}{
\begin{tabular}{l l c c c c c c}
\toprule
\textbf{Mission} 	  & \textbf{\mulrows{Organization\\(Country)}} & \textbf{\mulrows{Launch\\ year}} & \textbf{GNSS tracked} & \mulrows{\textbf{Number of} \\ \textbf{satellites}}   & \textbf{\mulrows{Receiving \\ channels}} & \textbf{Size} $[$cm$^3]$ & \textbf{Mass} $[$kg$]$\\
\midrule
\midrule
UK-DMC		  &  BNSC (UK)			& 2003 		  & GPS			& 1 &  1 	& 60$\times$60$\times$60  &$\sim$ 90\\
\midrule
TDS-1 		  & UKSA (UK)			& 2014 		  & GPS, Galileo & 1 &  4		& 77$\times$50$\times$90 & $\sim$ 157\\
\midrule
CYGNSS		  & 	NASA (USA)			& 2016 		  & GPS 			& 8 &  4		& 163.5$\times$52.1$\times$22.9 & $\sim$ 29\\
\midrule
Spire Batch-1  & \mulrows{Spire Global \\ (USA)} & 2019 & \mulrows{GPS, QZSS, \\ Galileo, SBAS} & 2 & $16-24$ & 10$\times$10$\times$30 & 5\\
\midrule
\textsuperscript{3}Cat-5/A & \mulrows{ESA (Europe) -\\ UPC (Spain)} & 2020	  & GPS, Galileo & 1 & 1 & 12$\times$24$\times$36 & $\sim$ 8\\
\midrule
PRETTY		  & \mulrows{RUAG GmbH \\ (Austria)}  & 2021 & GPS & 1 & - & 30$\times$10$\times$10  & $<$ 6\\
\midrule
GEROS-ISS 	  & GFZ	(Germany)	& 2022		  & \mulrows{GPS, Galileo, \\ GLONASS, BeiDou} & 1 & 4 & 117$\times$155$\times$86 & 376\\
\midrule
G-TERN 		  & \mulrows{GFZ (Germany) - \\ IEEC (Spain)} & 2025 & \mulrows{GPS, Galileo, \\ GLONASS, BeiDOu} & 1 & 12 & \mulrows{150$\times$110$\times$110 $-$ \\ 200$\times$120$\times$200} & 800 \\
\bottomrule 
\end{tabular}
}
\end{table}
The opportunity to deploy large constellations of satellites, combined with the availability of a large number of potential GNSS signal sources, allows achieving unprecedented revisit times on a global scale and paves the way for near-real time MS and EO \cite{camps2019nanosatellites}. As a matter of fact, the CYGNSS constellation offers a revisit time of 2.8 hour (median) and 7.2 hour (mean) between $38^\circ$ North and $38^\circ$ South latitudes \cite{ruf2017nasa}.
Nevertheless, revisit times offered by GNSS-R could be further decreased by increasing the number of satellites (massive GNSS-R constellations of more than 100 members might be  operational in the near future \cite{camps2019nanosatellites}) and/or by increasing the number of GNSS receiving channels mounted onboard each payload. It has been demonstrated, by means of simulation and statistical studies, that a mean revisit time down to 2.25 hours could be achieved with a constellation of 32 satellites, each equipped with at least 12 receiving channels and enabled to process GNSS signals from GPS, Galileo, GLONASS, and BeiDou-2~\cite{disimone2017sea,disimone2017gnss}. 
So far, primary remote sensing applications of spaceborne GNSS-R concern the analysis of the sea state (local wind speed, sea surface roughness, sea altimetry), soil moisture, biomass and vegetation estimation, sea-ice sheets analysis (height, volume, sea/ice index) \cite{jin2014gnss,zavorotny2014tutorial}. 
Different observables have been defined depending on the application. One of the most common and widely adopted, and from which other observables can be derived, is the delay-Doppler map (DDM). A DDM can be regarded as a 2D representation of the spatial distribution of the power scattered from the Earth' surface and exhibits the typical horseshoe shape over sea \cite{zavorotny2014tutorial}. Other common observables include the delay waveform, i.e., the zero-Doppler profile of the DDM, the DDM volume, both typically adopted for sea surface analysis, e.g., wind speed estimation; the carrier phase and the group delay lag between direct and reflected paths for sea altimetry; the reflectivity ratio for biomass and soil moisture estimation.
More recently, spaceborne GNSS-R is experiencing burgeoning interest for MS, as it appears to be very attractive solution. For instance, some works focus on using spaceborne GNSS-R data for oil slick detection~\cite{valencia2012using,li2014dual}, whereas application to ship detection is discussed for example in \cite{clarizia2015target,disimone2017sea,southwell2020matched}. 
%\cite{clarizia2015target}, \cite{southwell2020matched}. 
\begin{figure}[!t]
\centering
\includegraphics[width=0.5\columnwidth]{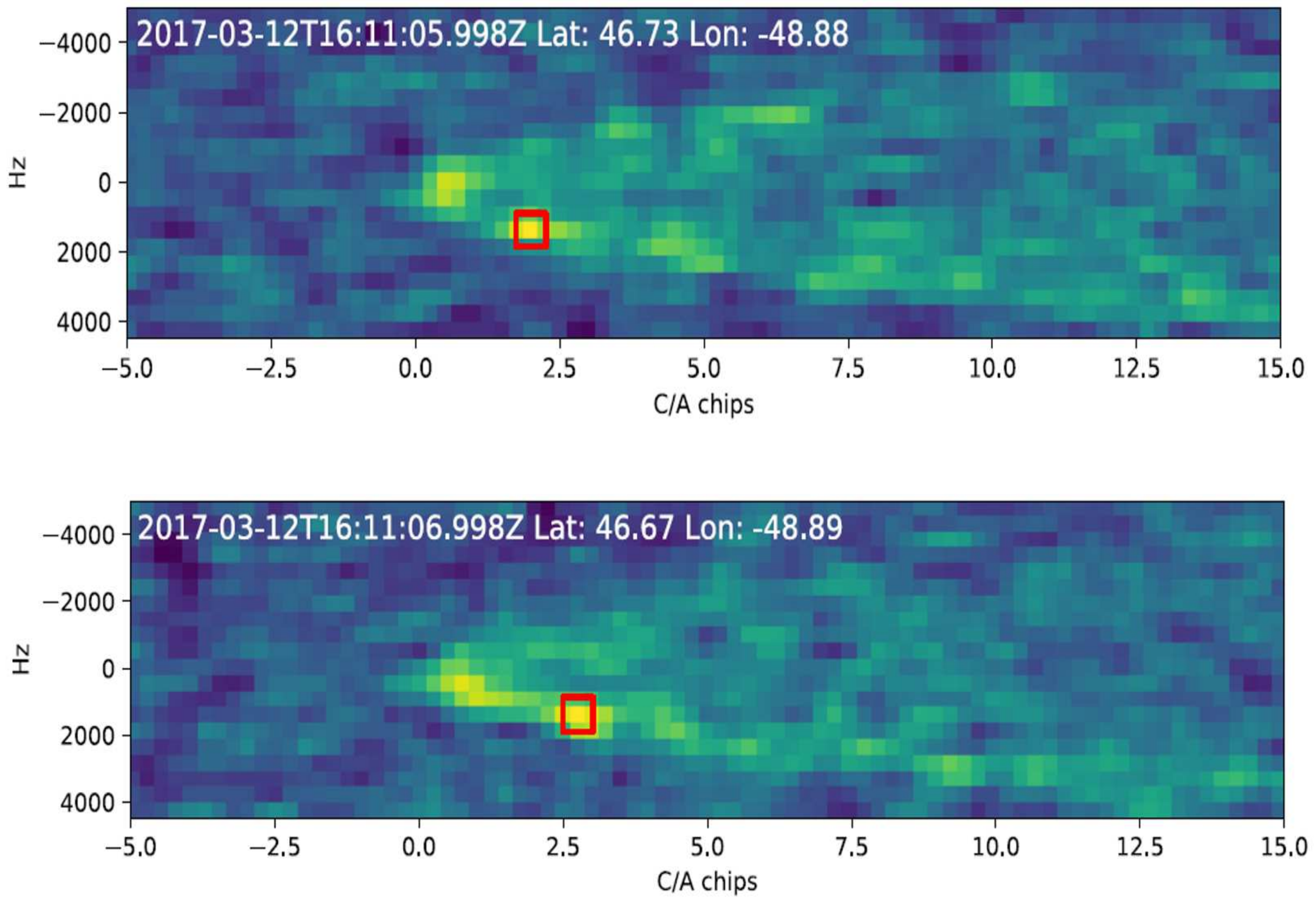}
\caption{Examples of spaceborne DDMs acquired by TechDemoSat-1 in the North Atlantic Ocean. Pixel spacing is 500 Hz in Doppler and 244 ns in delay. Both red square boxes highlight the target Hibernia platform located at 46.75009$^\circ$ N, 48.78161$^\circ$ W, which is visible as a bright feature.}
\label{fig:DDM examples}
\end{figure}
These studies have highlighted that applicability of standard GNSS-R to MS is mainly impaired by the poor spatial resolution - on the order of few kilometers close to the specular reflection point - and by the very low SNR of the signal reflected from even large ships. However, structures as large as oil platforms can still be detected in spaceborne DDMs acquired in open sea~\cite{disimone2017gnss}. Fig.~\ref{fig:DDM examples} shows two spaceborne DDMs acquired by TechDemoSat-1 in the North Atlantic Ocean. Apart the typical horseshoe shape exhibited by DDMs in open sea, a bright feature is visible as well (highlighted with red squares). It has been recognized as the Hibernia oil platform, located at 46.75009 N, 48.7816 W. Several factors have been found to cause poor visibility of ship targets within spaceborne GNSS-R DDMs. First, GNSSs provide a power density as low as $3\times 10^{-14} \text{ W}/\text{m}^2$ at the Earth' surface \cite{he2005}, thus motivating the strong temporal multilook typically performed by GNSS-R instruments \cite{zavorotny2014tutorial}.  Additionally, as illustrated in recent studies~\cite{disimone2018spaceborne,disimoneRCS,beltramonte2020simulation}, standard GNSS-R is poorly suited to ship detection applications for the following reasons: in the presence of ship targets, multiple-bounce scattering mechanisms arise from the EM coupling between sea surface and ship hull. Such scattering contributions enhance the energy reflected in the backscattering direction, which is not captured by standard GNSS-R receivers. Moreover, ships with metallic hulls reverse the polarization of the impinging EM wave. Accordingly, the backscattered ship echo is mainly RHCP, whereas a small fraction is conveyed in LHCP and collected by standard GNSS-R. Such considerations have been confirmed via theoretical and simulation link budget analyses~\cite{disimone2018spaceborne,beltramonte2020simulation}. In these works, the benefits of using non-standard GNSS-R instruments operating in backscattering configuration and sensitive to the RHCP channel have been highlighted in the context of MS. Further improvements of detection performance might be achieved by applying different processing chains to GNSS-R raw data, such as deconvolution techniques and bistatic SAR \cite{valencia2012using,antoniou2013gnss}. %\cite{antoniou2013gnss}.
To summarize, despite the intriguing features, e.g., passivity, exploitation of multi-frequency/multi-angle data, reliability of signal source, very low revisit time worldwide, offered by space-based GNSS-R for MS compared to other EO technologies, the feasibility of spaceborne GNSS-R data for MS applications, e.g., ship detection/tracking, requires further research.

\begin{table*}\caption{Comparison of various space-based technologies for MS applications}
\centering
\resizebox{.95\columnwidth}{!}{
\label{table:overview_comp}
\begin{tabular}{lll}
\toprule
\textbf{Technology} & \textbf{Advantages} & \textbf{Drawbacks} \\
\midrule
\midrule
\multirow{4}*{Sat-AIS} 	& Accurate ship information & Vulnerable (e.g. spoofing)\\
						& Very high update rate (up to 2 seconds) & Limited to ships with AIS\\
						& Global coverage of AIS services & Limited to cooperative ships\\
						& Compact, low-power, light-weight and cheap & Limited to ship detection/tracking\\
\midrule
\multirow{4}*{SAR} 	& All-weather, all-time sensing capabilities & Large size\\
					& Very high spatial resolution (down to 1 m) & \mulrows{Imagery difficult to visually \\ and manually interpret}\\
					& Polarimetric diversity & Affected by speckle noise\\
\midrule
\multirow{4}*{MSP}  &  Imagery easy to interpret & \mulrows{Sensitive to cloud and\\ sunlight conditions}\\
& \mulrows{Very high spatial resolution \\ (down to 0.3 m)} &  Limited revisit time \\
&  	& \mulrows{Limited areas covered during \\ each acquisition} \\
\midrule
\multirow{3}*{HSP} & Very high spectral resolution (down to 1 nm) & Low spatial resolution \\
& Anti-camouflage capabilities &  Large computational burden\\
& Suited to accurate classification & \mulrows{Sensitive to cloud and \\ sunlight conditions}\\
\midrule
\multirow{2}*{GNSS-R} & All-weather, all-time sensing capabilities & Low spatial resolution\\
& Compact, low-power, light-weight and cheap & Very low power density\\
& Seamless global coverage & \mulrows{Poor performance in standard \\ configuration}\\
& 100+ GNSS satellites available & Predetermined waveform\\
\bottomrule
\end{tabular}
}
\end{table*}

A summary comparison of the various remote sensing systems analyzed in this paper is provided in Table~\ref{table:overview_comp}.

\section{Conclusions}
\label{sec:conclusion_fut_work}
 
Maritime surveillance (MS) is crucial for search and rescue operations, fishery monitoring, pollution control, law enforcement, and national security policies. The goal of MS
is to provide 
%% seamless 
wide-area operational pictures of ship traffic
in real time. Most of the terrestrial sensors, i.e., terrestrial radars or the ground-based automatic identification system (AIS), are not able to guarantee a seamless coverage in remote ares of the oceans. However, with the increase in satellite constellations MS even in inaccessible areas becomes possible.
The first part of this work provided an overview of the main space-based sensor technologies, i.e., satellite AIS (Sat-AIS), synthetic aperture radar (SAR), multi-spectral (MSP) and hyper-spectral (HSP) optical sensors and global navigation satellite reflectometry (GNSS-R), and presented the advantages and disadvantages of each technology when employed in the scope of MS.

Note that, the availability of multiple space-based sensors, providing images of the current maritime situational picture at different spectral and spatial resolutions, requires also the development of advanced artificial intelligence and data fusion algorithms. These techniques have to cope with the processing and fusion of valuable information acquired from multiple sensors and will be covered in the second part of this work~\cite{SpaceAESM_2:J20}.
%for target detection, segmentation and classification. The most recent deep-learning techniques have been reported and their effectiveness has been demonstrated in four use cases using SAR, high-resolution (HR) and very high-resolution (VHR) images. 
%The second part of the paper has instead focused on the most recent Bayesian and statistical techniques to extract valuable knowledge from the huge amount of collected historical Sat-AIS data, such as most common maritime routes, and for multitarget tracking (MTT) using information collected by heterogeneous space-based and terrestrial sensors. MTT algorithms based on the sum-product algorithm (SPA) have gained strong popularity thanks to their abilities for fusing information from different heterogeneous sources, for their scalability, i.e., low computational complexity at the increasing number of information sources, targets and measurements, for their capability at including contextual information, such as maritime routes and class information. 
%A use case scenario that demonstrate the fusion of  measurements extracted from SAR images an area near Malta island together with the Sat-AIS measurements highlights the effectiveness of the SPA-based MTT methods in the context of MSA using space-based sensors. 

\bibliographystyle{IEEEtran}
%\bibliography{Bibliography/IEEEabrv,Bibliography/WGroup,Bibliography/BiblioCV,Bibliography/Temp,Bibliography/Temp_AS,Bibliography/Temp_NF,Bibliography/myBib}
\bibliography{Space-Based-MSA-AESM-V4_Part_1.bbl}
%\bibliography{../Bibliography/IEEEabrv,../Bibliography/WGroup,../Bibliography/BiblioCV,../Bibliography/Temp}
%\bibliography{./Bibliography/IEEEabrv,./Bibliography/Wgroup,./Bibliography/BiblioCV,./Bibliography/Temp}
\end{document}